\documentclass[12pt]{article}

\usepackage{amssymb,amsmath,amsfonts,eurosym,geometry,ulem,graphicx,subcaption,color,setspace,sectsty,comment,footmisc,caption,pdflscape,array,hyperref,soul,booktabs,threeparttable,adjustbox,longtable,float,placeins,authblk}
\usepackage[authoryear]{natbib}
\newcolumntype{L}[1]{>{\raggedright\let\newline\\arraybackslash\hspace{0pt}}m{#1}}
\newcolumntype{C}[1]{>{\centering\let\newline\\arraybackslash\hspace{0pt}}m{#1}}
\newcolumntype{R}[1]{>{\raggedleft\let\newline\\arraybackslash\hspace{0pt}}m{#1}}

\begin{document}

\begin{titlepage}
\title{Competitive effects of transmission constraints in the German electricity market}

\author[a]{Alice Lixuan Xu\thanks{Corresponding author: alicelixuan@hotmail.com}}
\author[a]{Clemens Stiewe}
\affil[a]{\small \textit{Centre for Sustainability, Hertie School, Berlin}}
\date{June 30, 2026}
\maketitle
\begin{abstract}
\noindent This paper estimates the effect of cross-border transmission constraints on suspected market power abuse in the German wholesale electricity market. Using a 2SRI instrumental variables approach, we study suspected strategic behavior by German gas- and coal-fired power plants in 2022-2024. Cross-border transmission constraints are measured using the maximum and minimum bounds of zonal net position, while suspected market power abuse is measured as the upward or downward deviation of observed dispatch from a modeled competitive benchmark. We find that transmission constraints significantly elevate the likelihood of suspected market power abuse. When headroom for further imports is already scarce, reducing import headroom by one Gigawatt (GW) increases the odds of suspected capacity withholding by 15\%. Similarly, reducing export headroom by one GW when it is scarce increases the odds of suspected capacity push-in, a strategy to depress prices, by 16\%. These results provide empirical support for interconnection expansion as an instrument to mitigate market power.   \\
\vspace{0in}
\vspace{0in}

\bigskip
\end{abstract}
\setcounter{page}{0}
\thispagestyle{empty}
\end{titlepage}
\pagebreak \newpage

\doublespacing

\section{Introduction} \label{sec:introduction}
Two fundamental characteristics of electricity markets -- grid constraints and short-run inelastic demand \citep{wilsonArchitecturePowerMarkets2002, hirthHowAggregateElectricity2024} -- make them vulnerable to market power abuse. Generators can have substantial time- and location-specific market power when competition is limited by local grid constraints \citep{wolakMeasuringUnilateralMarket2003}.
In turn, cross-border trade capacity mitigates market power by increasing competition from outside the local market \citep{borensteinMarketPowerElectricity1999}. Both the mitigating effect of increased transmission capacity \citep{borensteinCompetitiveEffectsTransmission2000} and the exacerbating effect of transmission congestion on market power abuse \citep{nappuMarketPowerImplication2013} have been shown in simulation studies, while empirical evidence remains scarce. 
\par
Using 2022-2024 unit-level data from the German wholesale electricity market, this paper empirically tests whether transmission constraints increase the likelihood of suspected market power abuse, extending the literature with a first estimate of the competitive effects of transmission congestion using publicly available data.
\par
Our results corroborate that transmission constraints exacerbate suspected strategic behavior: when the headroom for additional imports is low, reducing import headroom by one GW increases the odds of suspected capacity withholding by 15\%. Generators with market power may also benefit from strategically pushing capacity into the market to reduce prices if they are over-hedged on forward markets. We find that a one GW reduction in export headroom increases the odds of generators pushing capacity into the market 16\% when the headroom for additional exports is low. 
\par
The rest of this paper is organized as follows. \autoref{sec:literature} reviews the literature and sets out the theoretical framework of this analysis. \autoref{sec:empirical_strategy} explains our empirical strategy and introduces the econometric model setup, as well as data. \autoref{sec:result} presents our results, while \autoref{sec:discussion} discusses their limitations and implications. \autoref{sec:conclusion} concludes.

\section{Literature} \label{sec:literature}
A prominent form of market power abuse in wholesale electricity markets is capacity withholding. By reducing the capacity offered to the market, a generation company may be able to increase the clearing price and thus the inframarginal rent earned by its remaining capacity \citep{joskowQuantitativeAnalysisPricing2002}. A generator that has sold some of production forward has a mitigated incentive to withhold capacity because less of its remaining capacity is exposed to spot market prices \citep{wolakEmpiricalAnalysisImpact2000}. Meanwhile, \citeauthor{wolakEmpiricalAnalysisImpact2000} also shows that if a generator is overhedged (i.e., it has sold more forward than it produces), it may have an incentive to produce at prices below marginal cost, i.e., to push capacity into the market at negative margins. Opposite to capacity withholding, which aims to increase prices, generating at negative margins strategically lowers prices. Capacity push-in is therefore profitable if the gains from settling a short forward position at lower spot prices outweigh the costs of selling on the spot market at negative margins. As company hedge rates are high in the German electricity market \citep{uniperUniperCapitalMarkets2022,frankeRWEOffsetsFalling14/05/2020-15:17:00}, we study both capacity withholding and push-in.
\par
Residual demand curves are an established conceptual tool for market power studies in electricity markets (\citealp{borensteinCompetitiveEffectsTransmission2000, wolakEmpiricalAnalysisImpact2000,borensteinMeasuringMarketInefficiencies2002,hortacsuUnderstandingStrategicBidding2008}; \citealp[p.~309]{zotero-item-4460}). Residual demand is the difference between total market demand and the supply of all other companies, and measures the aggregate demand an individual company faces. The slope of the residual demand curve describes how readily other market participants would adjust their output in response to one company's strategic shift in output. In electricity markets, it summarizes the willingness and ability of other generators and loads to adjust their generation or consumption.
\par
\begin{figure}
    \centering
    \includegraphics[width=0.8\textwidth,trim={0cm 7.5cm 0cm 7cm},clip]{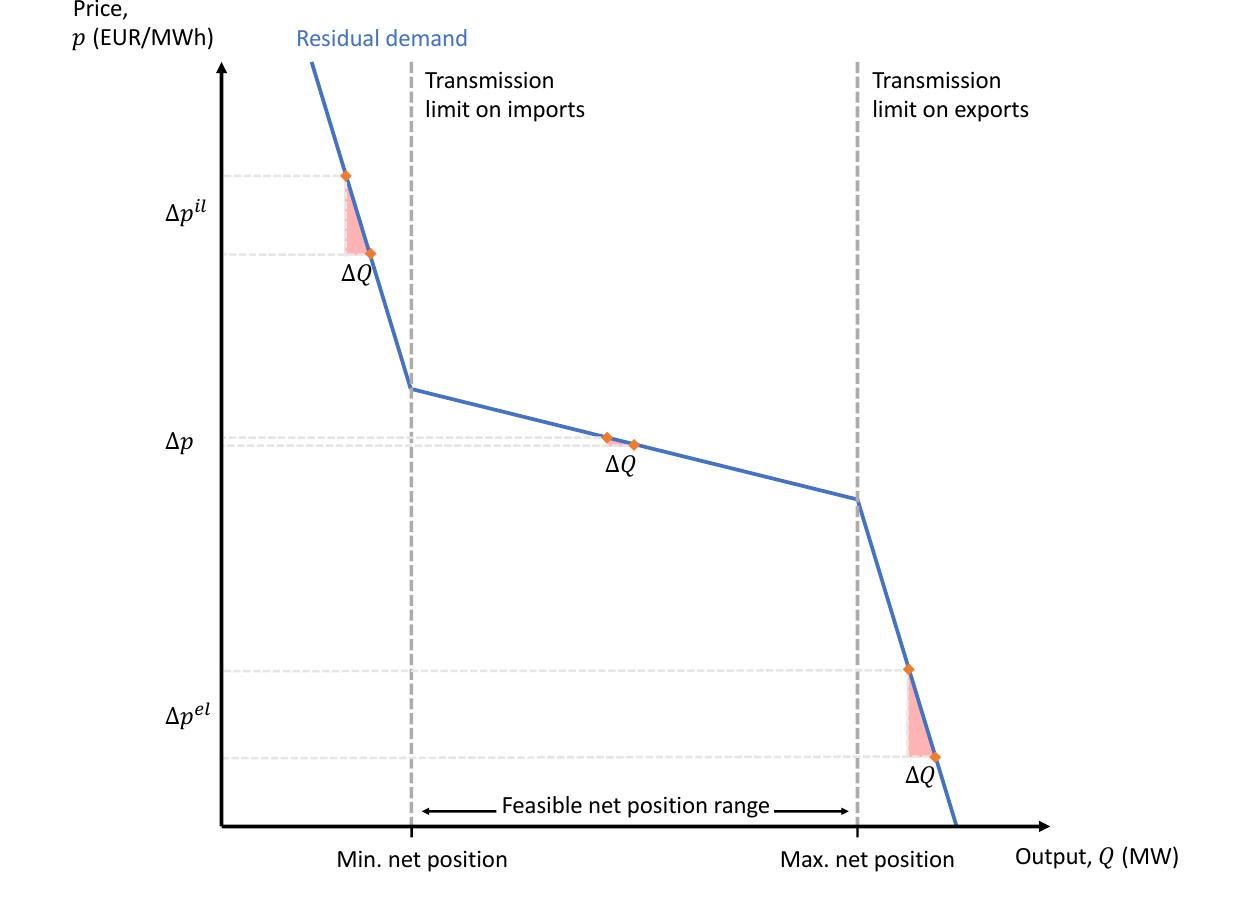}
    \caption{Residual demand under binding and non-binding transmission constraints. Net position denotes net export for the local market in European flow-based market coupling. Minimum and maximum net position denote the largest feasible imports and exports, respectively.} 
    \label{fig:residual_demand}
\end{figure}

\autoref{fig:residual_demand} shows how the elasticity of residual demand changes when import and export constraints become binding. A marginal change in output $\Delta Q$ can induce a larger price change ($\Delta p^{il}$ and $\Delta p^{el}$) than $\Delta p$ when the local market's import or export potential is constrained. This is because transmission constraints effectively limit the pool of competitors (i.e., generators and loads in neighboring markets) for local generators. Notably, withholding exacerbates congestion in the import direction but risks relieving it in the export direction, making it a more attractive strategy under import than under export constraints. The opposite holds for capacity push-in, which risks alleviating import constraints but exacerbates exports constraints. We therefore study the effect of import constraints on suspected capacity withholding and that of export constraints on suspected capacity push-in. 
\par
Detecting market power abuse in wholesale electricity markets is not straightforward and empirical studies on the topic have used various indicators. While some focus on aggregate outcomes, such as price, others focus on granular indicators of market participants' behavior, such as bid quantities.
\par
Price is a common indicator for market power abuse. Spreads between realized prices and competitive benchmarks \citep{borensteinMeasuringMarketInefficiencies2002, kwokaPriceSpikesEnergy2011, fladungElectricityPricesEnergy2024}, as well as price-cost margins measured with the Lerner Index\footnote{Lerner Index is defined as (Price - Marginal cost) / Price.} \citep{wolakMeasuringUnilateralMarket2003,pullerPricingFirmConduct2007,mulderCompetitionDutchElectricity2015} are established in the literature. Meanwhile, \citet{pullerPricingFirmConduct2007} points out that higher price-cost margins in the 2000 Californian electricity crisis can be either attributed to companies behaving less competitively or to companies behaving competitively but facing less elastic demand. This highlights that aggregate price-based indicators may struggle to detect uncompetitive behavior. More granular indicators can overcome that challenge and pin down specific forms of market power abuse.
\par
Capacity withholding has been measured at the firm level via lower-than-competitive generation output \citep{joskowQuantitativeAnalysisPricing2002}, diverging bid patterns between dominant and fringe players \citep{wolframStrategicBiddingMultiunit1998,itoSequentialMarketsMarket2016a}, and outage reports associated with high price periods \citep{berglerStrategicCapacityWithholding2017,fogelbergStrategicWithholdingProduction2019a, durmazGenerationFailuresStrategic2024}. \cite{batailleScreeningInstrumentsMonitoring2019} propose the Return on Withholding Capacity index, which emphasizes the underlying economic incentives. \cite{xuMarketPowerAbuse2025} show how the profitability of uncompetitive strategies drives the deviation of unit-level dispatch from a competitive benchmark, in both negative (withholding) and positive (push-in) directions. We build on their detection framework to measure market power abuse in this study and ask a research question that is upstream of the mechanism in \citeauthor{xuMarketPowerAbuse2025}. We aim to identify the effect of one specific driver of profitability -- transmission constraints -- on power plant generation behavior.
\par
Despite the well-established theoretical foundations, there is little empirical research on the competitive effects of transmission constraints. This may be explained by the complexity of measuring market power abuse, the unavailability of granular data for many markets, and the challenge of identifying exogenous variation in cross-border trade. Notably, \citet{ryanCompetitiveEffectsTransmission2021a} studies bids in the Indian electricity market and finds that markups are higher during congested hours. Using hourly unit-level bid, generation, price, and congestion data, \citet{wolakMeasuringCompetitivenessBenefits2015} quantifies the competitive benefits of transmission expansion in the Alberta wholesale electricity market. 
\par
In this study, we build on previous literature to examine the empirical relationship between suspected market power abuse and cross-border congestion in Germany in 2022-2024. We contribute to the literature in three ways: We provide a first causal estimate of the competitive effects of electricity cross-border trade using publicly available data. Second, we distinguish how import and export constraints drive different forms of strategic behavior. Lastly, we construct a novel metric for measuring transmission constraints in line with European flow-based market coupling (FBMC) methodology.

\section{Empirical strategy} \label{sec:empirical_strategy}
In this section, we introduce the treatment (transmission constraints) and outcome (market power abuse) variables, argue for our identification strategy, and describe the data used for this analysis.

\subsection{Measuring transmission constraints} \label{sec:transmission}
Import and export congestion reduce the elasticity of residual demand and are relevant inputs for market participants that bid strategically. Whether transmission constraints are binding is only determined during electricity market clearing and thus only known after market participants determine their (strategic) bids. We therefore approximate the likelihood that market participants assign to cross-border congestion, using data that are publicly available before market clearing.
\begin{figure}[htbp]
\centering
\includegraphics[width=0.45\textwidth]{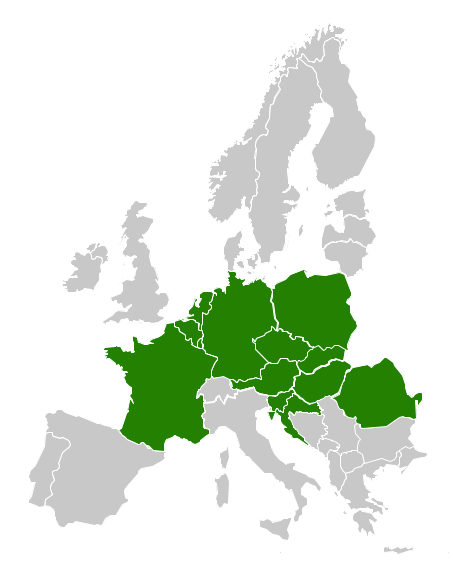}
\caption{Flow-based market coupling Core region participating markets: Austria, Belgium, Croatia, Czech Republic, France, Germany and Luxembourg, Hungary, Netherlands, Poland, Romania, Slovakia, Slovenia \citep{jointallocationofficeJointAllocationOffice2025}.}
\label{fig:map}
\end{figure}
\FloatBarrier
\par
\noindent
The flow-based domain is the central construct of FBMC and translates physical grid constraints into a feasible set of imports and exports per bidding zone \citep{schonheitFundamentalUnderstandingFlowbased2021}. It is calculated by the Coordinated Capacity Calculator (CCC) during the capacity calculation process that starts two days before delivery, and describes the feasible space of zonal net positions (i.e. zonal net exports, or zonal generation minus load) for all zones in the Core region. The 12 bidding zones that are part of the Core FMBC region are shown in \autoref{fig:map}. One day before delivery, during the day-ahead market clearing, the algorithm EUPHEMIA matches seller and buyer bids to maximize social welfare, subject to the constraints of the flow-based domain. Maximum and minimum net positions, which describe the largest feasible exports and imports per bidding zone, can be inferred from the flow-based domain and are published before the day-ahead market closes. These data allow market participants to predict how much headroom for cross-border trade will remain between zonal net position and its bounds at delivery.  
\par
\begin{figure}[htbp]
    \centering
    \includegraphics[width=0.95\textwidth]{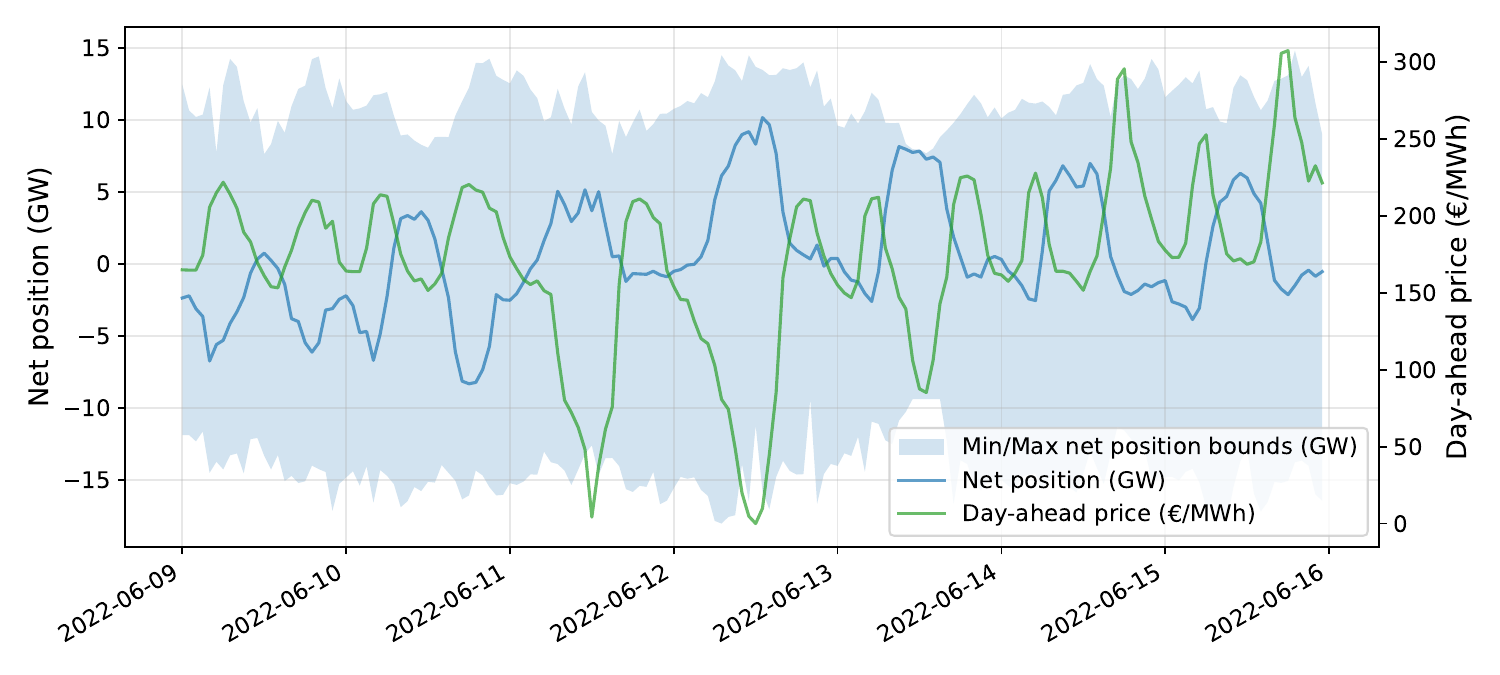}
    \caption{Core net position of the German-Luxembourgish market zone within maximum and minimum bounds, as well as day-ahead electricity prices, during the first week after the introduction of flow-based market coupling in the Core region in 2022.}
    \label{fig:np_price}
\end{figure}
\begin{figure}[htbp]
    \centering
    \includegraphics[width=0.65\textwidth]{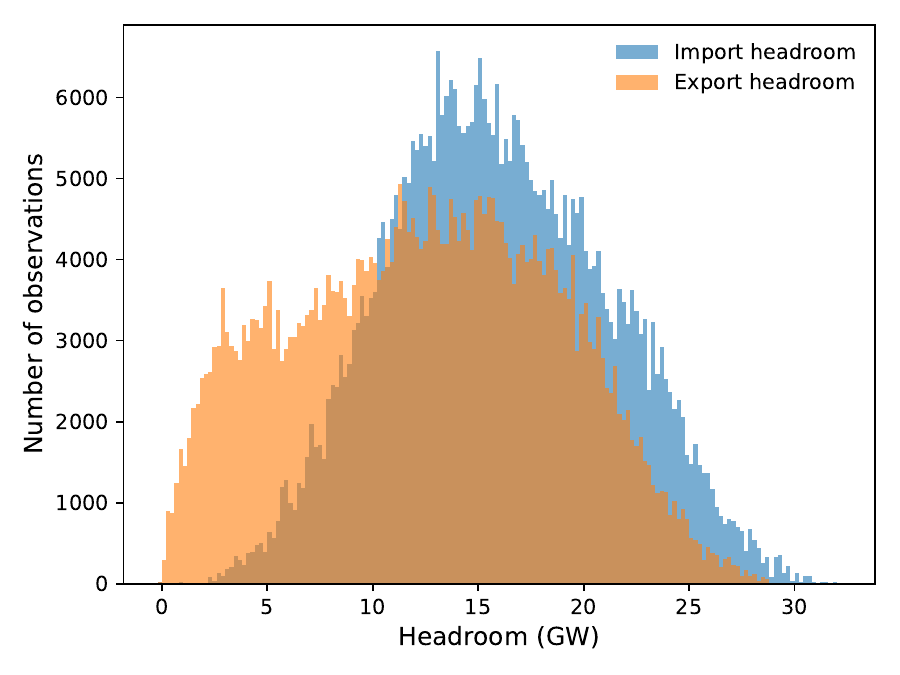}
    \caption{Histogram of hourly import and export headroom for the German-Luxembourgish market zone within the Core region, 2022-2024. Headroom is measured as the absolute value of the Gigawatt difference between maximum (for export) or minimum (for import) and realized net position in the Core region. Bin width: 200 MW.}
    \label{fig:headroom}
\end{figure}

We define import and export \textit{headroom} as the absolute difference (in GW) between the realized zonal net position in Core and the corresponding maximum (for export headroom) and minimum (for import) net position in Core. Headroom, in either import or export direction, hence indicates how much cross-border trade potential would remain for a given bidding zone per time step $t$. Headroom is specified as follows:
\begin{equation}
\begin{aligned}
H^{im}_t &= {NP}_t - {NP}^{min}_t, \\
H^{ex}_t &= {NP}^{max}_t - {NP}_t
\end{aligned}
\end{equation}
where ${NP}_t$ denotes the final Core net position, $ {NP}^{max}_t$ and ${NP}^{min}_t$ denote the bounds on maximum and minimum net positions (i.e., net exports). \autoref{fig:np_price} shows the dynamics of day-ahead prices and net position within its maximum and minimum limits for the German market zone during the first week of Core FMBC in 2022, with a clear negative correlation between prices and net exports. \autoref{fig:headroom} shows the histograms of import and export headroom for the German-Luxembourgish market during the sample period of 2022-2024. Export constraints appear to be binding more often than imports constraints, as shown by the many hours with export headroom close to zero. The different shapes of import and export headroom distributions may partly be explained by the data we use.

Our approach relies on using minimum and maximum net position data, which are only published for the Core FBMC region. Trade with neighbors of Germany that have not been part of Core during 2022-2024 -- Denmark and Switzerland -- is therefore not taken into account. Since Germany has been a net importer of electricity both from Denmark and Switzerland during our sample period \citep{SMARDEnergiedatenKompakt}, we likely overestimate import headroom and underestimate export headroom. This may explain why import headroom appears to be binding less often than export headroom in \autoref{fig:headroom}.

\subsection{Measuring market power abuse} \label{sec:mpa}
\begin{figure}[H]
\centering
\includegraphics[
    width=1\textwidth,
    trim={4cm 5cm 4cm 4.5cm},
    clip]{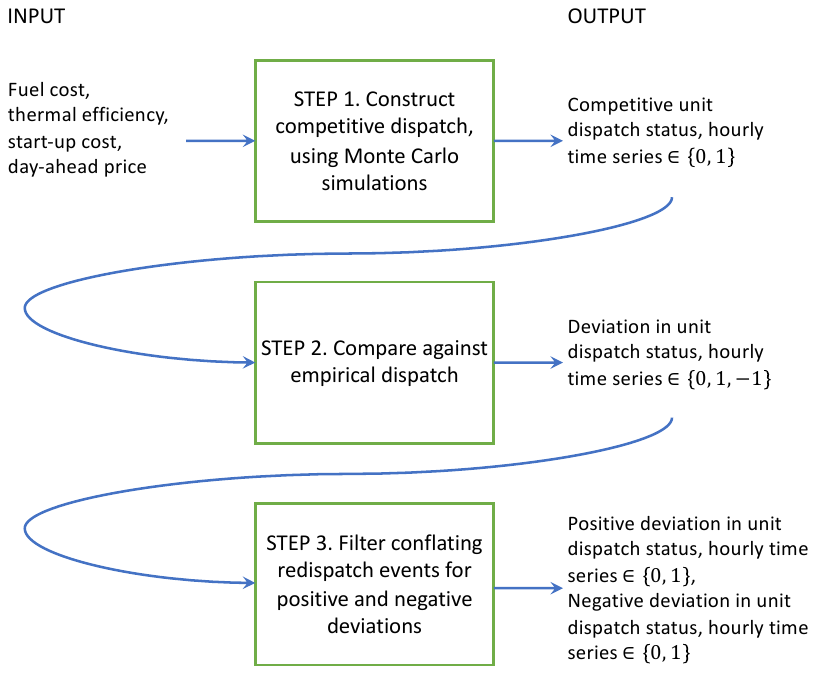}
\caption{Input-output data flow chart showing the 3-step construction of the outcome variable, deviation in dispatch as suspected market power abuse.}
\label{fig:method}
\end{figure}

We identify suspected market power abuse by comparing a generation unit's actual dispatch status to a counterfactual dispatch status estimated under competitive assumptions. This approach builds on \citet{xuMarketPowerAbuse2025} and consists of three steps, which we illustrate in \autoref{fig:method} and expand on the implementation in what follows.
\par
\noindent
We first construct a competitive dispatch time series for each generation unit in the sample, estimated based on their respective fuel costs, thermal conversion efficiencies, and cold start factors. The core assumption is that all units are price-takers and dispatch to maximize profits within each horizon, which spans 30 days plus one overlapping week per rolling window. Model uncertainty is accounted for by 1000 Monte Carlo iterations over key parameters -- unit-specific fuel costs, thermal conversion efficiency, and cold start factor -- such that 95\% of the sampled values land within $\pm 10\%$ variation around the mean in a normal distribution. A detailed documentation on the model setup for the competitive benchmark can be found in Appendix A of \citet{xuMarketPowerAbuse2025}. The resulting binary dispatch estimates of each iteration are then aggregated into a continuous index $\bar{d}_{it}$ between 0 and 1, such that
\begin{equation}
\bar{d}_{it} = \frac{1}{N}\sum_{n=1}^{N} \hat{d_{it}}, \qquad
\hat{d_{it}} \in \{0,1\}
\end{equation}
$\bar{d_{it}}$ can then be interpreted as the likelihood of unit $i$ at hour $t$ to be in dispatch under competitive assumptions, averaged across $N = 1000$ model runs. To ensure the confidence of the estimate, we discard hours where $\bar{d_{it}}$ is sensitive to the input parameters and keep only the hours where the simulation results are consistent for at least 95\% of the runs. This has left us with 79\% of the original sample. We code this information in $z_{it}$, such that it denotes the binary, estimated competitive counterfactual:
\begin{equation}
    z_{it} = 
    \begin{cases} 
    1, \qquad \bar{d_{it}} \ge 0.95 \\ 
    0, \qquad \bar{d_{it}} \le 0.05
    \end{cases}
\end{equation}
\noindent
At the second step, we compare actual unit-level generation status (i.e, generating or not generating) against its competitive estimate $z_{it}$. The resulting deviation in dispatch $y_{it}$ is then an indicator of when a unit's empirical dispatch deviates from expected dispatch under competitive assumptions. Three cases are differentiated: positive deviation, negative deviation, and no significant deviation in dispatch, such that:
\begin{equation}
y_{it} =
\begin{cases}
\quad1, \qquad \text{if } d_{it} =1 \text{ and } z_{it} =0 \\
\quad0, \qquad \text{if } d_{it} = z_{it} \\
\,-1, \qquad \text{if } d_{it} = 0 \text{ and } z_{it} =1
\end{cases}
\end{equation}
\noindent
where \( y_{it} \) denotes the deviation in dispatch of unit \( i \) at hour \( t \),\;$d_{it}$ the actual generation status, and $z_{it}$ the estimated competitive dispatch status.
\par
At the third step, we extend \citet{xuMarketPowerAbuse2025} by accounting for redispatch, which may require units to deviate from their market schedules. We adjust deviation markings based on generation-unit level redispatch event data as published by Netztransparenz. After identifying 15 out of the 25 generation units in our sample and assuming that the remaining 10 have not received redispatch orders in the sample period, we cleared deviation markings for suspected withholding if the downward redispatch order required more than 50\% of the unit's installed capacity. For suspected push-in, deviation markings were reverted to zero if the unit's actual generation in the relevant hour did not exceed the maximum power required by the upward redispatch.
\par
The resulting deviation in dispatch constitutes the binary outcome variable, $y_{it}$, of the following econometric model. Our final sample shows no significant deviations in 80\%, positive deviation in 12\%, and negative deviation 8\% of unit-hours. 
\par
Generation capacity can only be withheld if it was expected to be in dispatch in the first place, i.e., its marginal cost is lower than the day-ahead price. Similarly, capacity can only be pushed into the market if it was not expected to be in dispatch, i.e., its marginal cost is higher than the day-ahead price. In-the-money capacity can therefore only deviate negatively, while out-of-the-money capacity can only deviate positively. The outcome variable, $y_{it}$, is therefore binary, indicating whether a unit deviates negatively (when import constraints are binding) or positively (when export constraints are binding). 

\subsection{Econometric model}
\begin{figure}[htbp]
    \centering
    \includegraphics[width=0.9\textwidth,
        trim={1cm 9.5cm 1cm 9.5cm},
        clip]{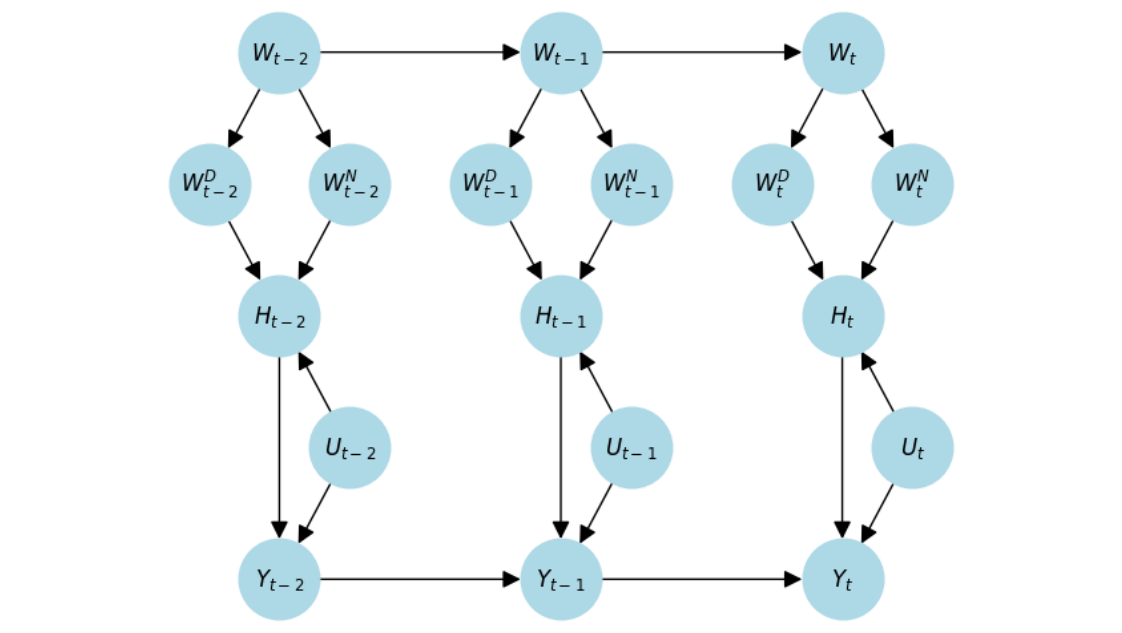}
    \caption{Directed acyclic graph for causal identification of the effect of headroom $H$ on deviation in dispatch $Y$. Wind is shown here as an example for wind,  solar, and load. $W^{D}$ stands for domestic and $W^{N}$ for neighboring wind generation, while $U$ encapsulates confounding variables.}
    \label{fig:DAG}
\end{figure}

This study aims to identify the causal effect of cross-border transmission constraints on the exercise of market power. The main challenges we face are reverse causality and autocorrelation, and we show how the two affect our causal identification in a directed acyclic graph (DAG)\footnote{A DAG illustrates the causal dependencies (edges) between variables (vertices), directed from each cause to its effect(s). A path following the directed edges depicts a causal mechanism that represents an understanding of the real world. For an introduction, see \citet{zotero-item-4406}.} in Figure~\ref{fig:DAG}. Here, the DAG shows three consecutive time steps and the causal paths between and within each one of them. Each past time step affects the future one in regards to wind pattern, $W$, which represents our instruments in this stylized example, and deviation in dispatch, which is our outcome variable, $Y$. The main effect we want to identify is illustrated by the edge directed from $H_{t}$ to $Y_{t}$. Because both congestion and generation are equilibrium market outcomes, we represent confounding variables and potential reverse causality in the DAG with $U_{t}$, as suggested by \citet{tiedemannIdentifyingElasticitiesAutocorrelated2024}.

\noindent
Reverse causality can be an issue because strategy may aggravate or even create congestion \citep{nappuMarketPowerImplication2013}. Naive estimates might therefore be biased. To retrieve the causal effect of congestion on suspected market power abuse we hence follow an instrumental variables (IV) approach. Since our dependent variable, hourly deviation in dispatch from the competitive benchmark, is binary, we follow a Two-stage Residual Inclusion approach (2SRI). 2SRI is a control function approach that allows us to model non-linear outcomes \citep{terzaTwostageResidualInclusion2008,wooldridgeControlFunctionMethods2015}. Instead of replacing the endogenous variable with its estimated values from the first stage in the second stage, as in Two-stage Least Squares (2SLS), 2SRI uses the residuals from the first stage -- which is estimated through Ordinary Least Squares (OLS) -- as an additional regressor in the second stage. This retains the original endogenous variable in the second stage while controlling for endogeneity through the first-stage residuals \citep{wooldridgeControlFunctionMethods2015}. A significant effect of the first-stage residuals in the second stage then indicates the presence of endogeneity.
\par
Good instrumental variables fulfill two criteria: they are correlated with the endogenous regressor (instrument relevance) and uncorrelated with the error term (instrument exogeneity) \citep{wooldridgeEconometricAnalysisCross2010}. While instrument relevance is assessed using the first-stage F-statistic following \citet{staigerInstrumentalVariablesRegression1997}, justifying instrument exogeneity requires a theoretical discussion.
\par
Our instruments for endogenous import/export headroom are wind and solar electricity generation, as well as load, in neighboring markets. Neighboring wind, solar, and load are defined as the weighted average of total national wind generation, solar generation, or system load (in GW) across Germany's direct neighbors. Weights are given by the bilateral interconnector capacities between Germany and its neighbors as estimated in \citet{stieweCrossborderCannibalizationSpillover2025}. \citet{beltramiBordersEstimatingMarginal2025} also use wind electricity in adjacent markets as an instrument for electricity cross-border trade. Weather-based instrumental variables have been popular in econometric research (see, e.g., \citet{angristInterpretationInstrumentalVariables2000}) because weather is arguably exogenous. We argue similarly that wind, solar, and load in markets neighboring Germany affect the import and export potential of the German market. At the same time, neighboring wind and solar generation are weather-driven and therefore not affected by market power abuse in the German electricity market. Following the literature, we assume that electricity demand is largely inelastic and hence unaffected by strategic behavior as well. 
\par
Instrument exogeneity more generally requires that the instrument is uncorrelated with the error term of the structural equation. One might argue that the same weather system determines both domestic wind and solar generation as well as wind and solar generation in neighboring markets, especially if markets are geographically small. As long as we control for domestic wind and solar, however, it is not part of the error term anymore and instrument exogeneity is restored. Hence, neighboring wind and solar only affect our outcome variable through their effect on import/export headroom. 
\par
The second main challenge we face is autocorrelation. Because of inert wind patterns and thermal ramping constraints, wind and thermal generation are both autocorrelated. This again threatens the exogeneity of our instrument. The issue can be clearly illustrated in \autoref{fig:DAG}, where $W_{t-1}$ affects our instrument $W^{N}_{t}$ via its autocorrelation with $W_{t}$. This violates exogeneity of the instrument because $W_{t-1}$ also affects the outcome $Y_{t}$ via $Y_{t-1}$. It is difficult, however, to get a meaningful measurement of wind patterns $W_{t}$ for entire regions. Hence, we can block the path from $W_{t-1}$ to our outcome and thus restore exogeneity either by conditioning on the lagged terms of domestic ($W^{D}_{t-1}$) and neighboring ($W^{N}_{t-1}$) wind generation or on lagged deviation in dispatch, $Y_{t-1}$. Since $Y_{t-1}$ is an estimate that carries measurement error, we condition on lagged instruments.
\par
Since electricity supply curves are usually convex, the functional form of the effect of headroom on strategic behavior is likely different for import and export constraints. The stylized residual demand curves in \autoref{fig:method} suggest a discrete change in elasticity, and thus in the likelihood of strategic behavior, as soon as interconnection is exhausted in either direction. With convex supply, however, residual demand should become steeper \textit{before} all constraints on critical network elements become binding at the minimum net position bound (i.e., at zero import headroom): When imports are high but some headroom still remains, the potential for imports from neighboring markets with cheaper generators is already exhausted (i.e., no margin remains on lines that import from these markets). Withholding capacity in this situation results in a larger price increase than at ample import headroom.
\par
Convex supply also implies that residual demand remains elastic before export constraints become binding. When export headroom is tight, pushing capacity into the market (e.g. by bidding zero marginal cost) displaces other low-marginal-cost generators, leading to a smaller price effect than at tight import headroom. Only when all network elements are constrained in the export direction at the maximum net position, the elasticity in residual demand sharply decreases. Taken together, we therefore expect that the slope of residual demand that German generators face is rather steep towards the import limit and smaller towards the export limit.
\par
We explore nonlinear effects of headroom by including both the linear and squared headroom terms in our regression. We estimate two separate models for positive and negative deviation in dispatch from the competitive baseline, regressing the probability of negative deviation (i.e., suspected capacity withholding) on import headroom and the probability of positive deviation (i.e., suspected capacity push-in) on export headroom. Both of these models are estimated with 2SRI. In the first stage of the 2SRI models, we estimate $H^{n}_{t}$ using OLS, with $n \in \{im,\, ex\}$. Headroom is a system-level quantity and the same for all units in hour $t$. We use $H^{n}_{t}$ in what follows and let $H^{n}_{it} \equiv H^{n}_{t}$ for every unit $i$.

\begin{equation}
\begin{aligned}
H^{n}_{t}
&= \delta_{0}
 + \underbrace{\delta_{1}\,wind^{nb}_{t}
 + \delta_{2}\,solar^{nb}_{t}
 + \delta_{3}\,load^{nb}_{t}}_{\text{instruments}} \\
&\quad
 + \delta_{4}\,wind_{t}
 + \delta_{5}\,solar_{t}
 + \delta_{6}\,load_{t} \\
&\quad
 + \delta_{7}\,wind^{nb}_{t-1}
 + \delta_{8}\,solar^{nb}_{t-1}
 + \delta_{9}\,load^{nb}_{t-1} \\
&\quad
 + \delta_{10}\,wind^{nb}_{t-2}
 + \delta_{11}\,solar^{nb}_{t-2}
 + \delta_{12}\,load^{nb}_{t-2} \\
&\quad
 + \delta_{13}\,wind_{t-1}
 + \delta_{14}\,solar_{t-1}
 + \delta_{15}\,load_{t-1} \\
&\quad
 + \delta_{16}\,wind_{t-2}
 + \delta_{17}\,solar_{t-2}
 + \delta_{18}\,load_{t-2} \\
&\quad
 + \delta_{19}\,carbon_{t}
 + \delta_{20}\,gas_{t}
 + \delta_{21}\,coal_{t} \\
&\quad
 + \alpha_i + \gamma_m + v_{it},
\end{aligned}
\end{equation}
\noindent
where neighboring wind ($wind^{nb}_{t}$), solar ($solar^{nb}_{t}$), and load ($load^{nb}_{t}$) are the instrumental variables. Domestic equivalents, $wind_{t}$, $solar_{t}$, and $load_{t}$, as well as the first and second lags of domestic and neighboring wind, solar, and load, are included as exogenous controls. Following \citet{canalesEmpiricalEstimateElectricity2025}, we include gas ($gas_{t}$), coal ($coal_{t}$), and carbon ($carbon_{t}$) prices, which affect both the shape of supply curves and headroom. This allows us to isolate the effect of headroom on strategic behavior. Entity fixed effects are denoted by $\alpha_i$ and month fixed effects by $\gamma_m$. We obtain the first-stage residuals:
\begin{equation}
\hat{v}_{it} = H^{n}_{t} - \widehat{H^{n}_{it}}.
\end{equation}
\noindent
where the fitted value $\widehat{H^{n}_{it}}$ is entity-specific because we included entity fixed effects. In the second stage, the likelihood that a generation unit deviates from the competitive baseline ($y_{it}=1$) is modeled as a logit. Following \citet{terzaTwostageResidualInclusion2008} and \citet{wooldridgeControlFunctionMethods2015}, we include the endogenous variable $H^{n}_{t}$, a squared term $(H^{n}_{t})^2$, and first-stage residuals $\hat{v}_{it}$:

\begin{equation}
\begin{aligned}
\Pr(y_{it}=1 \mid X_{it}, \alpha_i, \gamma_m)
&= \Lambda\Bigg(
  \beta_{0}
  + \beta_{1}\,H^{n}_{t}
  + \beta_{2}\,\left(H^{n}_{t}\right)^2 \\
&\quad
  + \rho\,\hat{v}_{it} \\
&\quad
  + \beta_{3}\,wind_{t}
  + \beta_{4}\,solar_{t}
  + \beta_{5}\,load_{t} \\
&\quad
  + \beta_{6}\,wind^{nb}_{t-1}
  + \beta_{7}\,solar^{nb}_{t-1}
  + \beta_{8}\,load^{nb}_{t-1} \\
&\quad
  + \beta_{9}\,wind^{nb}_{t-2}
  + \beta_{10}\,solar^{nb}_{t-2}
  + \beta_{11}\,load^{nb}_{t-2} \\
&\quad
  + \beta_{12}\,wind_{t-1}
  + \beta_{13}\,solar_{t-1}
  + \beta_{14}\,load_{t-1} \\
&\quad
  + \beta_{15}\,wind_{t-2}
  + \beta_{16}\,solar_{t-2}
  + \beta_{17}\,load_{t-2} \\
&\quad
  + \beta_{18}\,carbon_{t}
  + \beta_{19}\,gas_{t}
  + \beta_{20}\,coal_{t} \\
&\quad
  + \alpha_i + \gamma_m
\Bigg),
\end{aligned}
\end{equation}
\noindent
where $\Lambda(\cdot)$ denotes the logistic cumulative distribution function,  $\hat{v}_{it}$ is the first-stage residual, and $X_{it}$ collects headroom, squared headroom, first-stage residuals, and the controls. Since $\hat{v}_{it}$ is a generated regressor, analytic standard errors are inconsistent \citep{paganEconometricIssuesAnalysis1984}. We therefore report clustered bootstrap standard errors based on 500 iterations.
\par
\subsection{Data}
We obtain realized, minimum, and maximum Core net position data from the Joint Allocation Office (JAO), the central trading platform of TSOs, which publishes the data via their publication tool \citep{jointallocationofficeJointAllocationOffice2025}. Market data on day-ahead price, load, as well as wind and solar generation are obtained from the ENTSO-E transparency platform \citep{entso-eENTSOETransparencyPlatform2025}. We obtain hourly power plant dispatch and hourly available capacity data from the EEX transparency platform \citep{eexEEXTransparencyPlatform2024}, as preprocessed by \citet{fusarbassiniTimeSeriesAvailable2025}. For fuel prices, we use historical futures indices for Dutch TTF gas \citep{investing.comDutchTTFNatural2025} and CIF ARA coal \citep{investing.comCoalAPI2CIF2025}, as well the European Union Emissions Trading System (EU ETS) price \citep{AllowancePriceExplorer}. Summary statistics are shown in \autoref{tab:summary}.

\begin{table}[htbp]
\centering
\caption{Summary statistics}
\label{tab:summary}
\begin{tabular}{lccccccc}
\hline
 & Min & P25 & P50 & P75 & Mean & Std.\ Dev. & Max \\
\hline
Import headroom [GW] & 0.00 & 11.01 & 14.29 & 18.11 & 14.76 & 5.19 & 31.91 \\
Export headroom [GW] & 0.00 & 6.03 & 11.09 & 15.80 & 11.14 & 6.09 & 28.79 \\
Gas price [EUR/MWh] & 22.93 & 33.11 & 40.30 & 66.90 & 66.32 & 56.90 & 339.19 \\
Carbon price [EUR/t] & 52.51 & 68.08 & 76.51 & 85.57 & 76.66 & 10.27 & 98.01 \\
Coal price [USD/t] & 93.25 & 114.05 & 121.50 & 174.15 & 163.47 & 84.01 & 394.50 \\
Wind [GW] & 0.04 & 4.63 & 9.79 & 18.69 & 12.77 & 10.23 & 48.73 \\
Solar [GW] & 0.00 & 0.00 & 0.09 & 10.60 & 6.66 & 10.33 & 47.25 \\
Load [GW] & 30.85 & 46.21 & 53.32 & 61.34 & 53.67 & 9.19 & 73.91 \\
\hline
\end{tabular}
\end{table}

\section{Results} \label{sec:result}
For both models, we find that generators are significantly more likely to deviate from the competitive benchmark when headroom for cross-border trade is low. While the effect of lower import headroom on the likelihood of suspected capacity withholding is linear across the headroom distribution, the effect of lower export headroom on the likelihood of suspected capacity push-in is sizable only at low headroom levels. We describe our results in more detail below, starting with the negative deviation model.

\begin{figure}[H]
    \centering
    \includegraphics[width=\textwidth]{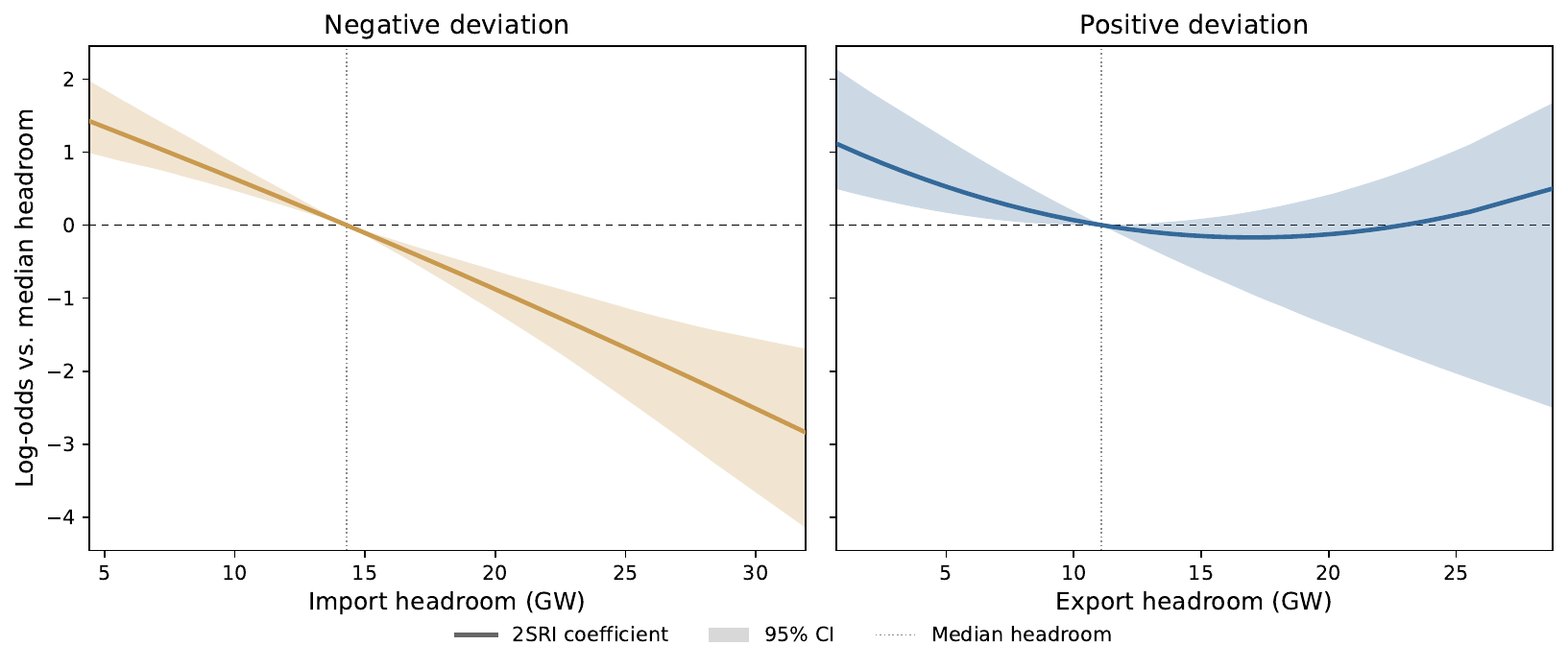}
    \caption{Predicted log-odds of strategic deviation as a function of headroom, relative to median headroom, for the 2SRI and plain logit specifications.}
    \label{fig:results}
\end{figure}
\noindent
Neighboring wind, solar, and load are relevant instruments for import headroom in our 2SRI model for suspected capacity withholding (see first-stage F-statistic in \autoref{tab:iv_2sri}). The second-stage headroom coefficients show that generators are significantly more likely to withhold capacity when headroom is low. The insignificant quadratic term for import headroom reveals that the effect is essentially linear. A one GW reduction in import headroom, evaluated at median import headroom, thus increases the odds of suspected capacity withholding by 16\%.\footnote{$\Delta\text{odds} ~(\%) = \left(e^{-(\beta_{1} + 2\beta_{2} H)} - 1\right)\times 100 = \left(e^{-(-0.133 + 2\times(-0.0006)\times 14.3)} - 1\right)\times 100 \approx 16\%$, where $\beta_{1} = -0.133$ is the coefficient on $H^{im}_{t}$ and $\beta_{2} = -0.0006$ the coefficient on $(H^{im}_{t})^{2}$, evaluated at the median import headroom of $14.3$ GW.} At the 5th percentile (ca. 7 GW), i.e., when the potential for additional imports is low, the change in odds is 15\% and hence virtually identical.

Our second 2SRI model also exhibits a strong first stage and finds that lower export headroom significantly increases the probability of suspected capacity push-in. Here, the effect is non-linear and shrinks towards zero at large export headroom. A one GW reduction in export headroom increases the odds of suspected capacity push-in by 6\% at median export headroom (ca. 11 GW), rising to 16\% at the 5th percentile (ca. 2 GW). \autoref{fig:results} shows how the log-odds of suspected capacity withholding and push-in change across the import and export headroom distributions: In both models, the likelihood of suspected strategic deviation from the modeled competitive benchmark increases as marginal trade potential approaches zero. For the negative deviation model, this holds even when headroom is above median. The positive deviation model, however, yields a non-linear effect which disappears as export headroom reaches its median value. In \autoref{sec:functional}, we investigate the shape of the effect, using a more flexible functional form.
\par
The size of the effect is comparable across the two models. Reducing import headroom by one GW at the bottom 5th percentile of headroom increases the probability of negative deviation by 3.5 percentage points (p.p.) for a unit with baseline deviation probability of 50\%. The same reduction in export headroom increases the probability of positive deviation by 4.0 p.p.
\par
First-stage residuals $\hat{v}_{it}$ are significant only in the model for negative deviation, suggesting the presence of endogeneity. Because $\hat{v}_{it}$ has a positive sign for imports, a plain logit model would likely yield a less negative effect of import headroom on suspected capacity withholding. For the positive deviation model, a plain logit might be a more efficient -- and unbiased -- estimator than 2SRI. We investigate endogeneity in the following.

\begin{table}[h]
\centering
\begin{adjustbox}{max width=\textwidth}
\begin{threeparttable}
\caption{2SRI regression results}
\label{tab:iv_2sri}
\begin{tabular}{lcc}
\toprule
 & Negative deviation & Positive deviation \\
\midrule
Intercept & 3.57** & 1.85 \\
 & (1.78) & (2.09) \\
\addlinespace[0.5em]
$H^{ex}_{t}$ &  & -0.16*** \\
 &  & (0.04) \\
\addlinespace[0.5em]
$(H^{ex}_{t})^{2}$ &  & 0.00*** \\
 &  & (0.00) \\
\addlinespace[0.5em]
$H^{im}_{t}$ & -0.13*** &  \\
 & (0.05) &  \\
\addlinespace[0.5em]
$(H^{im}_{t})^{2}$ & -0.00 &  \\
 & (0.00) &  \\
\addlinespace[0.5em]
$\hat{v}_{it}$ & 0.13*** & 0.02 \\
 & (0.03) & (0.04) \\
\addlinespace[0.5em]
Entity and month FE & Yes & Yes \\
Controls & Yes & Yes \\
\midrule
Observations & 173365 & 227213 \\
Pseudo $R^{2}$ & 0.278 & 0.471 \\
\midrule
First-stage joint $F$-stat & 4350.0 & 4432.5 \\
First-stage $R^{2}$ & 0.564 & 0.710 \\
\bottomrule
\end{tabular}
\begin{tablenotes}[flushleft]
\item Clustered bootstrap standard errors in parentheses.
\item Controls include contemporaneous values and lags of wind, solar, and load, lags of neighboring wind, solar, and load, as well as carbon, gas, and coal prices. Full coefficient estimates are reported in Table~\ref{tab:iv_2sri_appendix}.
\item * p $<$ 0.10, ** p $<$ 0.05, *** p $<$ 0.01
\end{tablenotes}
\end{threeparttable}
\end{adjustbox}
\end{table}
\FloatBarrier
\subsection{Investigating endogeneity} \label{sec:endogeneity}

To assess the size and sign of the endogeneity bias, we compare our 2SRI IV results with a plain logit model that resembles the second stage of our 2SRI model. For suspected capacity withholding, the plain logit -- unlike the IV model -- finds an insignificant effect of import headroom. The effect of export headroom on suspected capacity push-in, on the other hand, is similar for the plain logit and 2SRI models. This aligns with the significant and negative effect of $\hat{v}_{it}$ in the negative deviation model and an insignificant effect in the positive deviation model. It confirms that endogeneity only affects the negative deviation model and shows that the effect of export headroom on positive deviation from the competitive benchmark could be estimated more efficiently with a plain logit model.

\begin{figure}[H]
    \centering
    \includegraphics[width=\textwidth]{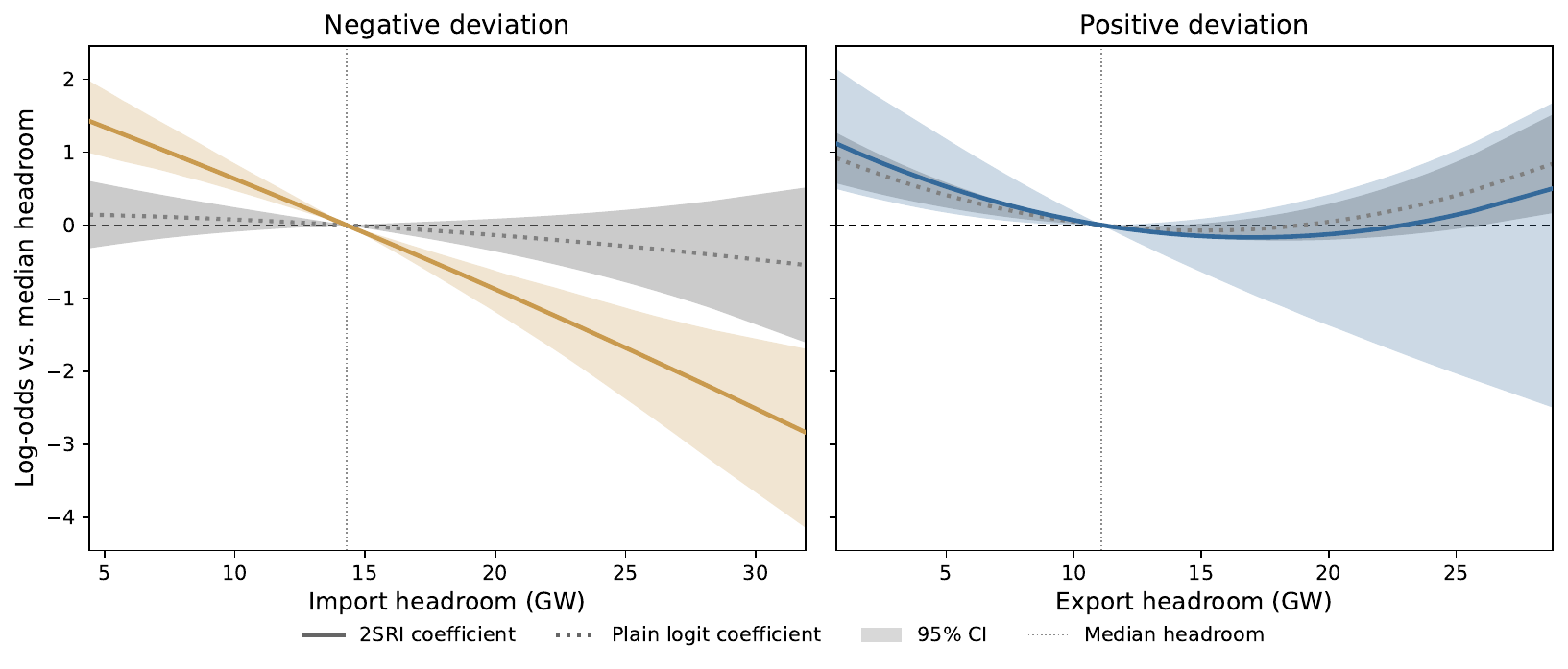}
    \caption{Predicted log-odds of strategic deviation as a function of headroom, relative to median headroom, for the 2SRI and plain logit specifications.}
    \label{fig:results_plain}
\end{figure}
\noindent
The sign of the endogeneity bias for the negative deviation model does not suggest meaningful reverse causality: Withholding capacity during import-constrained hours would exacerbate import constraints and therefore yield a stronger association between headroom and suspected strategic behavior, not a weaker one. It is, however, consistent with endogeneity from measurement error in the treatment variable, which biases the coefficients on the mismeasured variable towards zero \citep{wooldridgeEconometricAnalysisCross2010}. In \autoref{sec:transmission} we discussed the measurement error in headroom due to the omission of German cross-border flows with Denmark and Switzerland in our data. Omitting sizable imports (mostly from Denmark) implies overestimating import headroom when it is already low, which means that export headroom should be underestimated when it is large. Since the effect of headroom should concentrate at low headroom, measurement error should have a more pronounced effect in the negative deviation model. This is consistent with the attenuated effect of headroom we find in the plain logit model for negative deviation.
\par
Unlike the headroom variable, our instrumental variables -- neighboring wind, solar, and load -- are calculated including Denmark and Switzerland. The 2SRI IV estimates therefore better capture the reduced elasticity of residual demand that German thermal generators face when German import headroom is low.

\subsection{Investigating functional form} \label{sec:functional}
To investigate whether our baseline models capture the functional form of the effect of headroom well, we run a robustness check that uses a more flexible specification. Instead of modeling the effect of linear and squared headroom terms, we assign observations to decile bins of the headroom distribution and construct a dummy variable for each bin, omitting the 40--50th percentile bin as the reference category.
\par
\begin{figure}[H] 
    \centering
    \includegraphics[width=\textwidth]{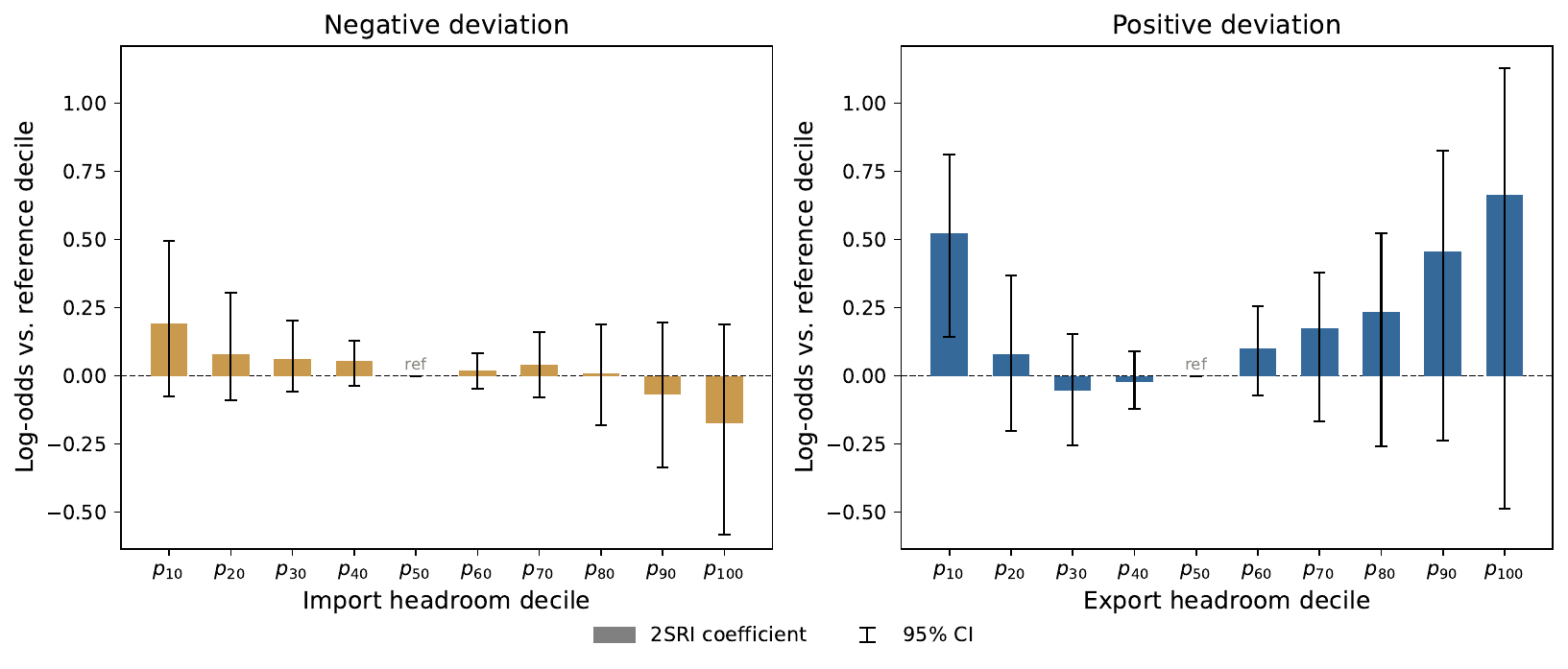}
    \caption{2SRI headroom decile coefficients. Bars show log-odds differences relative to the omitted median headroom decile ($H_t^{p50}$).}
    \label{fig:bin_result}
\end{figure}

As in the baseline models, the flexible specification reveals an approximately linear shape of the effect on negative deviation and a non-linear effect of export headroom  (see \autoref{fig:bin_result}). However, due to reduced precision with fewer observations in each bin, none of the headroom decile dummy variables is significant in the negative deviation model. In the positive deviation model, however, the lowest headroom decile is associated with 69\% higher odds of suspected capacity push-in relative to the median decile ($\exp(0.52) - 1 \approx +69\%$). Full regression results for the flexible specification can be found in the appendix in \autoref{tab:iv_2sri_bin}. A more linear effect in the negative deviation model and a more discrete effect in the positive deviation model is in line with the intuition that the withholding strategy faces a more gradually increasing residual demand elasticity as more import constraints become binding, while the push-in strategy faces a step change in residual demand elasticity when all export constraints become binding.

\section{Discussion} \label{sec:discussion}

This study shows that transmission congestion is associated with a higher incidence of suspected strategic behavior by coal- and gas-fired power plants in the German electricity market. While the theoretical foundation for this effect -- reduced competition when transmission constraints are binding -- is established in the literature, the empirical identification of the effect is not trivial.
\par
This is because detecting market power abuse is notoriously difficult in electricity markets. The approach we have chosen critically relies on the accuracy of the competitive benchmark scenario. False positives (i.e., flagging suspected market power abuse while there has actually been none) can result from over- or underestimating marginal costs. If, for example, a unit's marginal costs were underestimated, it may appear to be withholding capacity while its true marginal costs are above spot prices. Overestimation of marginal costs may lead to false positives for capacity push-in. To reduce the number of false positives, we follow the conservative approach of \citet{xuMarketPowerAbuse2025} that only flags any unit-hour as suspicious when at least 95\% of the Monte Carlo iterations agree on the competitive generation status. Monte Carlo simulations therefore mitigate but cannot rule out false positives: they are still possible if the chosen cost parameters are inaccurate or if the uncertainty bands around these parameters are too narrow. 
\par
Discretizing deviation, i.e., only flagging hours as suspected market power abuse when a unit is fully idle, is another measure we take to reduce the number of false positives. This, however, leaves any partial withholding undetected and illustrates that a strategy to reduce false positives would naturally increase the number of false negatives, i.e., classifying unit-hours as competitive even if market power abuse took place. 
\par
False positives and false negatives constitute misclassification of the binary outcome \citep{hausmanMisclassificationDependentVariable1998}. \citeauthor{hausmanMisclassificationDependentVariable1998} show that misclassification biases estimated coefficients on independent variables towards zero if the probability of misclassification does not depend on the covariates, whereas the direction of the bias is ambiguous if misclassification varies with the covariates (i.e., with headroom). Two scenarios are relevant for this study: A higher likelihood for false negatives at low headroom is acceptable because it biases headroom coefficients towards zero.\footnote{Not capturing partial withholding is one example for that. Partial withholding should concentrate at low headroom, where the incentive is strongest as residual demand is less elastic. False negatives from partial withholding should therefore cluster where the effect is largest, attenuating the estimated headroom coefficient.} What is more problematic is an elevated likelihood of false positives at low headroom, which inflates headroom coefficients. While our Monte Carlo approach generally leans towards under-reporting market power abuse, it does not prevent a clustering of false positives at low headroom. 
\par
Consistently underestimating the marginal cost of peaking plants, which often operate when imports are high (and import headroom is low) is one potential source of such clustering. Similarly, overestimating the marginal cost of lignite plants, which set the price more often at low export headroom, would inflate the effect of export headroom in the positive deviation model. Balancing market participation, which is not captured in our model due to the lack of public data, might result in false positives too: A unit providing non-spinning, stand-by reserve might be wrongly flagged as strategically withholding capacity if their modeled marginal costs are below day-ahead prices, whereas units providing spinning reserve might be missclassified as strategically pushing in capacity if their modeled marginal costs exceed day-ahead prices. However, it is not straightforward that balancing services are relatively more attractive at low import/export headroom: \citet{justPricingReservesValuing2008} show that when contracting periods are long, the incentives to provide spinning reserves are not affected by spot price levels, which correlate with import and export headroom.
\par
A separate question is whether our results are driven by a few outlier units. We address this by conducting a leave-one-out analysis, which is summarized in \autoref{fig:leaveoneout}. Omitting any of the 25 units in our sample does not move the marginal effect of headroom outside of the full-sample confidence interval for both negative and positive deviation, indicating that neither result is driven by any individual unit. The small differences in effect size may reflect genuinely stronger or weaker strategic behavior, or unit-specific misclassification.
\par
The models for negative and positive deviation from the competitive baseline are specified symmetrically, yet the two kinds of strategic behavior they try to detect differ in one crucial aspect, which is the hedge rate of the generator. A high hedge rate represents high opportunity costs for capacity withholding and makes capacity push-in more attractive. It is therefore reasonable to assume that generation firms specialize in either withholding or push-in strategies within their forward contract's settlement period.
\par
Unit-level negative and positive deviation rates are indeed negatively correlated (see \autoref{fig:entity_specialization}). This figure also shows that most of this specialization might be explained by the unit's production type. Specifically, combined cycle gas plants are more likely to withhold capacity because their foregone profits when withheld are lower than that of coal-fired plants, which typically have lower marginal costs and hence higher margins. On the other hand, lignite plants appear to specialize in push-in strategies, which may be explained by their higher foregone profits from withholding. Meanwhile, some heterogeneity within production type is apparent in \autoref{fig:entity_specialization}, which may be explained by different hedge rates across companies. Finally, the implausibly high average deviation rates for some units in \autoref{fig:entity_specialization} suggest possible false positives despite our efforts to reduce them. However, \autoref{fig:leaveoneout} does not indicate that the misclassification of individual units drives the overall result: omitting the units with the highest deviation rates leaves the headroom coefficients essentially unchanged. Nevertheless, this cannot rule out potential bias from misclassification. Constructing a more accurate competitive benchmark is a promising direction for future research on identifying strategic behavior.

\section{Conclusion} \label{sec:conclusion}
This paper examines whether cross-border trade constraints affect strategic dispatch decisions in the German electricity market. Using 2022-2024 data, we find that lower headroom for imports and exports significantly increases the probability of fossil generators to withhold or push in capacity, respectively. While withholding capacity from the market increases prices and thus allows remaining capacity to capture higher inframarginal rent, the economic rationale for capacity push-in is that the gains from settling forward contracts at lower spot prices outweigh the costs of selling on the spot market at prices below marginal costs. Such market power abuse is more profitable when transmission constraints are binding, because they make the residual demand faced by market participants less elastic.
\par
The effect of import headroom on suspected capacity withholding is approximately linear across the headroom distribution, while the effect of export headroom on suspected capacity push-in is substantial at low headroom levels only. This suggests that the shape of residual demand, which is defined as total demand minus supply of competitors, reflects that electricity supply curves are often convex. With a convex supply curve, the slope of residual demand should be increasing as import headroom is reduced while being flat before export constraints become binding due to zero-marginal-cost renewable generation.
\par
Market power abuse is difficult to measure in electricity markets, and our approach cannot fully rule out that competitive generation patterns are flagged as strategic. Distinguishing strategic from competitive behavior is an open challenge for future research. Despite these limitations, our central finding is robust: deviations consistent with capacity withholding and push-in become more frequent precisely when the incentive to behave strategically, driven by low headroom for cross-border trade, is strongest.
\par
The policy implication of this study is straightforward: Expanding interconnection, both by building new transmission lines and by using existing capacity more efficiently, mitigates market power abuse.

\clearpage

\subsection*{Authorship contribution statement}
Both authors contributed equally to this work.

\subsection*{Data availability}
The code and data needed to replicate the analysis are available under an open license and can be found at https://github.com/alicelixuan/MPA-transmission

\subsection*{Declaration of use of generative AI and AI-assisted technologies}
During the preparation of this work, the authors used Claude to assist with writing the code for the analysis. The authors take full responsibility for the content of the published article.

\subsection*{Acknowledgement}
We thank Lion Hirth, Jorge Sánchez Canales, Chiara Fusar Bassini, Frank Boerman-Lima, Mark A. Kayser, Will Lowe, and participants at the Young Energy Economists and Engineers Seminar for valuable feedback and discussions. This work is supported by the German Federal Ministry of Research, Technology and Space (BMFTR) via the ARIADNE Project (FKZ 03SFK5K0-2).

\clearpage
\appendix

\renewcommand{\thetable}{A\arabic{table}}
\renewcommand{\thefigure}{A\arabic{figure}}
\setcounter{table}{0}
\setcounter{figure}{0}

\section*{Appendix} \label{sec:appendix}

\begin{table}[]
\centering
\begin{adjustbox}{max width=\textwidth}
\begin{threeparttable}
\caption{Full 2SRI regression results}
\label{tab:iv_2sri_appendix}
\begin{tabular}{lcc}
\toprule
 & Negative deviation & Positive deviation \\
\midrule
Intercept & 3.57** (1.78) & 1.85 (2.09) \\
\addlinespace[0.3em]
$H^{ex}_{t}$ &  & -0.16*** (0.04) \\
\addlinespace[0.3em]
$(H^{ex}_{t})^{2}$ &  & 0.00*** (0.00) \\
\addlinespace[0.3em]
$H^{im}_{t}$ & -0.13*** (0.05) &  \\
\addlinespace[0.3em]
$(H^{im}_{t})^{2}$ & -0.00 (0.00) &  \\
\addlinespace[0.3em]
$\hat{v}_{it}$ & 0.13*** (0.03) & 0.02 (0.04) \\
\addlinespace[0.3em]
$wind_{t}$ & 0.13*** (0.01) & -0.03 (0.03) \\
\addlinespace[0.3em]
$solar_{t}$ & 0.12*** (0.02) & -0.05*** (0.02) \\
\addlinespace[0.3em]
$load_{t}$ & -0.14*** (0.01) & 0.05*** (0.02) \\
\addlinespace[0.3em]
$wind_{t-1}$ & -0.01** (0.01) & 0.01* (0.01) \\
\addlinespace[0.3em]
$solar_{t-1}$ & -0.04*** (0.01) & 0.03 (0.02) \\
\addlinespace[0.3em]
$load_{t-1}$ & 0.05*** (0.01) & -0.02** (0.01) \\
\addlinespace[0.3em]
$wind_{t-2}$ & 0.00 (0.01) & -0.00 (0.01) \\
\addlinespace[0.3em]
$solar_{t-2}$ & 0.00 (0.01) & -0.00 (0.01) \\
\addlinespace[0.3em]
$load_{t-2}$ & 0.00 (0.01) & 0.03 (0.02) \\
\addlinespace[0.3em]
$wind^{nb}_{t-1}$ & -0.11 (0.07) & -0.13** (0.06) \\
\addlinespace[0.3em]
$solar^{nb}_{t-1}$ & -0.03 (0.07) & 0.08 (0.11) \\
\addlinespace[0.3em]
$load^{nb}_{t-1}$ & -0.01 (0.03) & -0.08** (0.03) \\
\addlinespace[0.3em]
$wind^{nb}_{t-2}$ & 0.01 (0.04) & -0.01 (0.07) \\
\addlinespace[0.3em]
$solar^{nb}_{t-2}$ & -0.04 (0.09) & -0.03 (0.07) \\
\addlinespace[0.3em]
$load^{nb}_{t-2}$ & -0.06* (0.03) & -0.05 (0.07) \\
\addlinespace[0.3em]
$carbon_{t}$ & 0.01 (0.02) & -0.05* (0.03) \\
\addlinespace[0.3em]
$gas_{t}$ & 0.00 (0.00) & -0.00 (0.01) \\
\addlinespace[0.3em]
$coal_{t}$ & 0.00 (0.00) & -0.01 (0.00) \\
\addlinespace[0.3em]
Entity and month FE & Yes & Yes \\
\midrule
Observations & 173365 & 227213 \\
Pseudo $R^{2}$ & 0.278 & 0.471 \\
\midrule
First-stage joint $F$-stat & 4350.0 & 4432.5 \\
First-stage $R^{2}$ & 0.564 & 0.710 \\
\bottomrule
\end{tabular}
\begin{tablenotes}[flushleft]
\item Clustered bootstrap standard errors in parentheses.
\item * p $<$ 0.10, ** p $<$ 0.05, *** p $<$ 0.01
\end{tablenotes}
\end{threeparttable}
\end{adjustbox}
\end{table}

\begin{table}[]
\centering
\begin{adjustbox}{max width=\textwidth}
\begin{threeparttable}
\caption{2SRI regression results: decile bin specification}
\label{tab:iv_2sri_bin}
\begin{tabular}{lcc}
\toprule
 & Negative deviation & Positive deviation \\
\midrule
Intercept & 1.63 & -1.07 \\
 & (1.82) & (2.69) \\
\addlinespace[0.5em]
$H_{t}^{p10}$ & 0.19 & 0.52*** \\
 & (0.14) & (0.16) \\
\addlinespace[0.5em]
$H_{t}^{p20}$ & 0.08 & 0.08 \\
 & (0.11) & (0.14) \\
\addlinespace[0.5em]
$H_{t}^{p30}$ & 0.06 & -0.05 \\
 & (0.07) & (0.10) \\
\addlinespace[0.5em]
$H_{t}^{p40}$ & 0.06 & -0.02 \\
 & (0.04) & (0.06) \\
\addlinespace[0.5em]
$H_{t}^{p60}$ & 0.02 & 0.10 \\
 & (0.03) & (0.08) \\
\addlinespace[0.5em]
$H_{t}^{p70}$ & 0.04 & 0.17 \\
 & (0.06) & (0.14) \\
\addlinespace[0.5em]
$H_{t}^{p80}$ & 0.01 & 0.24 \\
 & (0.09) & (0.20) \\
\addlinespace[0.5em]
$H_{t}^{p90}$ & -0.07 & 0.46* \\
 & (0.14) & (0.27) \\
\addlinespace[0.5em]
$H_{t}^{p100}$ & -0.17 & 0.66* \\
 & (0.20) & (0.40) \\
\addlinespace[0.5em]
$\hat{v}_{it}$ & -0.00 & -0.05** \\
 & (0.03) & (0.02) \\
\addlinespace[0.5em]
Entity and month FE & Yes & Yes \\
Controls & Yes & Yes \\
\midrule
Observations & 173365 & 227213 \\
Pseudo $R^{2}$ & 0.277 & 0.471 \\
\midrule
First-stage joint $F$-stat & 4350.0 & 4432.5 \\
First-stage $R^{2}$ & 0.564 & 0.710 \\
\bottomrule
\end{tabular}
\begin{tablenotes}[flushleft]
\item Clustered bootstrap standard errors in parentheses.
\item * p $<$ 0.10, ** p $<$ 0.05, *** p $<$ 0.01
\end{tablenotes}
\end{threeparttable}
\end{adjustbox}
\end{table}

\begin{figure}[H] 
    \centering
    \includegraphics[width=0.45\textwidth]{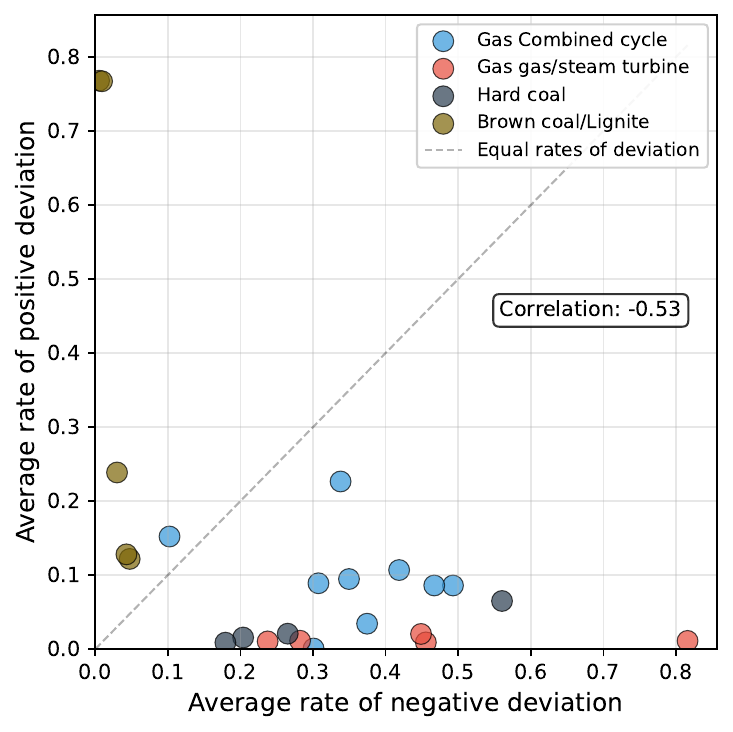}
    \caption{Average rates of negative and positive deviation from the competitive benchmark at the unit level. Each point represents one generator.}
    \label{fig:entity_specialization}
\end{figure}

\begin{figure}[H] 
    \centering
    \includegraphics[width=\textwidth]{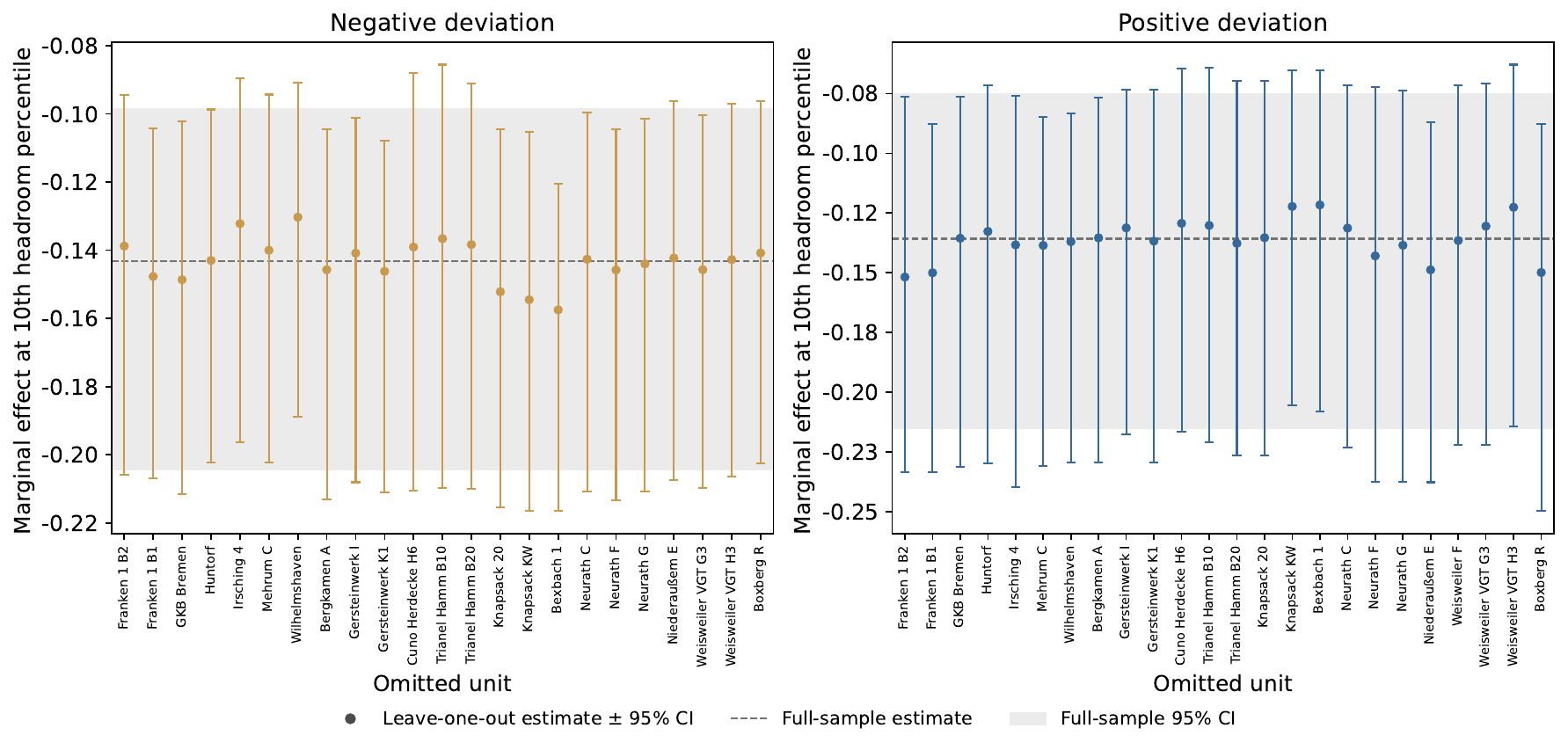}
    \caption{Leave-one-out sensitivity analysis for the marginal effect of headroom, evaluated at the 10th percentile of headroom.}
    \label{fig:leaveoneout}
\end{figure}

\clearpage
\bibliographystyle{apalike}  
\bibliography{xb_market_power}       

\end{document}